\documentclass[twoside]{article}
\usepackage{qic,epsfig}
\usepackage{graphicx}

\def\be{\begin{eqnarray}}
\def\ee{\end{eqnarray}}
\begin{document}
\setlength{\textheight}{8.0truein}    %FOR 2ND PAGE ONWARDS

\runninghead{Sufficient condition on noise correlations for scalable quantum computing}
            {J. Preskill}

\normalsize\textlineskip
\thispagestyle{empty}
\setcounter{page}{1}
%%% Hide copyright in preprint
%\copyrightheading{Vol.}{No.}{Year}{Page Nos.}
%%%

\vspace*{0.88truein}

\alphfootnote

\fpage{1}

\centerline{\bf
%%%%%%%%%%%%%%%%%%%%%
%Put in titles here
%%%%%%%%%%%%%%%%%%%%%
SUFFICIENT CONDITION ON NOISE CORRELATIONS}
\centerline{\bf FOR SCALABLE QUANTUM COMPUTING}
\vspace*{0.37truein}
\centerline{\footnotesize
%%%%%%%%%%%%%%%%%%%%%%%%%%%%%%%%%%%%
%put authors' name and address here
%%%%%%%%%%%%%%%%%%%%%%%%%%%%%%%%%%%%
JOHN PRESKILL}
\vspace*{0.015truein}
\centerline{\footnotesize\it Institute for Quantum Information and Matter, California Institute of Technology }
\baselineskip=10pt
\centerline{\footnotesize\it Pasadena, CA 91125, USA}
\vspace*{0.225truein}

%%% Hide dates in preprint
%\publisher{(received date)}{(revised date)}
%%%
%
%
\vspace*{0.21truein}
%
%% \abstracts{first paragraph}{second paragraph}{third paragraph}
%% If there is only one paragraph, just keep the second and third empty 
%% like the following one 
\abstracts{
%%%%%%%%%%%%%%%%%%%%
% put abstract here
%%%%%%%%%%%%%%%%%%%%
%%
I study the effectiveness of fault-tolerant quantum computation against correlated Hamiltonian noise, and derive a sufficient condition for scalability. Arbitrarily long quantum computations can be executed reliably provided that noise terms acting collectively on $k$ system qubits are sufficiently weak, and decay sufficiently rapidly with increasing $k$ and with increasing spatial separation of the qubits.
}{}{}
\vspace*{10pt}

\vspace*{3pt}
%%% Hide communicate in preprint
%\communicate{to be filled by the Editorial}
%%%
%%
%%
\vspace*{1pt}\textlineskip    %) USE THIS MEASUREMENT WHEN THERE IS
   %) A SECTION HEADING
%\vspace*{-0.5pt}
%\noindent
%%%%%%%%%%%%%%%%%%%%%%%%%%%%%%%%
%put the text of the paper here
%%%%%%%%%%%%%%%%%%%%%%%%%%%%%%%%%\documentclass[aps,pra,showpacs,preprint]{revtex4}

\section{Introduction}
\label{sec:intro}

Our planet is in the midst of a digital revolution, validating the scalability of classical information processing. Will quantum computers likewise be scalable, eventually performing tasks that surpass what could be done if the world were classical?

The accuracy threshold theorem for quantum computation establishes that scalability is achievable provided that the currently accepted principles of quantum physics hold and that the noise afflicting a quantum computer is neither too strong nor too strongly correlated  \cite{ben-or,kitaev_threshold,knill, jp_threshold,gottesman_threshold,AGP06,reichardt-threshold}. For scalability to fail as a matter of principle then, either quantum mechanics must fail for complex highly entangled systems (as 't Hooft \cite{tH99} has suggested), or else either the Hamiltonian or the quantum state of the world must impose noise correlations that overwhelm fault-tolerant quantum protocols (as Alicki {\em et al.} \cite{AHHH02, Alicki06,Alicki09} and Kalai \cite{Kalai05,Kalai06,Kalai08,Kalai09,Kalai11} have suggested). 

Because of the profound implications of large-scale quantum computing for computational complexity theory and fundamental physics, skepticism is natural and useful. Debate about the feasibility of fault-tolerant quantum computation can sharpen our understanding of the issues, and raise the stakes as quantum science and technology continue to advance. But skeptics should be pressed for a conception of Nature in which classical computing is feasible yet quantum computing is forbidden.

The theory of fault-tolerant quantum computing closely resembles the corresponding classical theory, but there are also important differences. Perhaps most fundamentally, a classical computer can perform reliably even if the information being processed leaks to the environment, but a quantum computation will fail unless the processed quantum information remains almost perfectly concealed. Indeed, in noisy classical systems expelling heat to the environment is essential to ensure controllability, while for quantum systems energy dissipation may induce decoherence and hence cause trouble.

A central lesson of fault-tolerant quantum computing is that this is a false dichotomy --- energy dissipation is just as crucial for reliable quantum computation as for classical computation \cite{aharonov_reversible}. The trick is to expel entropy without exposing the protected coherent quantum information to the environment. In principle this can be achieved using quantum error-correcting codes \cite{shor_qecc,steane_qecc} and carefully designed fault-tolerant protocols \cite{shor_ft,Gottesman09}. Another important lesson is that small rotation errors in imperfect unitary quantum gates can be digitized and hence corrected like the bit flip errors in a dissipative classical system.

The goal of this paper is to exhibit a class of noise models for which quantum computing is provably scalable. We will assume that qubits can be refreshed on demand, {\em i.e.}, that it is possible to prepare a standard initial state of $n$ qubits, approximating the product state $|0\rangle^{\otimes n}$, with small, weakly correlated errors. One might imagine that Nature conspires to block the creation of a good approximation to a pure product state, but such a limitation would threaten the scalability of classical computation as well, and so does not seem like a promising way to make a fundamental distinction between classical and quantum computation. 

More plausibly, the distinction might arise because noise correlations unavoidably obstruct the creation of profoundly entangled states of many qubits. We will address this issue by studying a class of Hamiltonian models of correlated noise and deriving a sufficient condition for scalability within the context of this class of models. Skeptics are invited to explain why no quantum engineer could ever build a system with noise meeting this criterion.

In our models, the noise correlations arise from an (in principle infinite) series of terms in the Hamiltonian, where terms acting on $k$ system qubits have an operator norm obeying an upper bound that drops sufficiently rapidly with increasing $k$ and with growing spatial separation among the qubits. We should emphasize that this Hamiltonian containing many-qubit terms is not meant to be the fundamental Hamiltonian describing the interactions among elementary particles and fields; rather it is the \emph{effective} coarse-grained Hamiltonian describing the residual interactions among qubits that have been ``dressed'' by integrating out high-frequency short-distance degrees of freedom. These residual interactions might be weak, and usefully characterized by an upper bound on the operator norm, even though the underlying interactions that dress the qubits are relatively strong. The qubits themselves might be quite complex composite objects.

Though our scalability criterion, formulated in terms of upper bounds on the norms of terms in this effective Hamiltonian, may be widely applicable, it would be desirable to relax the criterion in various ways. In particular, scalability can be proven for noise models in which system qubits couple to harmonic oscillator bath variables with unbounded norm, assuming the initial state of the bath meets certain conditions (for example, if the bath starts out in a low-temperature Gibbs state) \cite{NP09}. But, so far, a proof of scalability for a system in contact with an oscillator bath has been worked out only for Gaussian noise, {\em i.e.}, for the case where the noise is completely characterized by the two-point correlations function of the bath variables. Extending that argument to a nonlinear oscillator bath with non-Gaussian correlations remains a technically challenging open problem, worthy of further attention. 

Gaussian noise models are often regarded as reasonably realistic in physical settings where the system is weakly coupled to many environmental
degrees of freedom \cite{Caldeira83}. But even in the Gaussian case, arguments for scalability hold only under assumptions about the frequency spectrum of bath fluctuations \cite{Terhal05,Novais07,Novais08,NP09, Novais10}, and some skeptics have criticized these assumptions \cite{Dyakonov06,Alicki07,Hines08}. However these critics have not clearly identified any class of Gaussian noise models which would allow high-fidelity gates in few qubit systems while disallowing large-scale fault-tolerant quantum computing. Hence if their objections carry weight, we may encounter a barrier blocking further systematic improvements in quantum gate fidelity in the relatively near future. The question we are trying to address here is not whether noise will limit the reliability of small-scale quantum computers but rather whether large scale quantum computing might eventually fail, even though small scale quantum computers continue to improve.

Anyway, without further apologies, we will use a Hamiltonian model in which the noise strength can be characterized using the operator norm. We go beyond previous work \cite{AKP06} by investigating the effects of not just few-body correlations in the noise but also sufficiently weak many-body correlations. The goal is to get a clearer picture of how harmful such noise correlations could be. The type of model we study has a notable advantage --- we do not need to make any assumption about the initial state of the bath to derive useful results. We formulate the model and state the main result in Sec. 2, then prove it in Sec. 3. 

\section{Noise model and scalability criterion}
The noise model we consider is formulated by specifying a time-dependent Hamiltonian $H$ that governs the joint evolution of the system and the bath, which can be expressed as
\be
H = H_S + H_B + H_{SB};
\ee
here $H_S$ is the time-dependent Hamiltonian of the system that realizes an ideal quantum circuit, $H_B$ is the Hamiltonian of the bath, and $H_{SB}$, which describes the coupling of the system to the bath, is the origin of the noise. We place no restrictions on the bath Hamiltonian $H_B$. Without any loss of generality, we may expand the system-bath Hamiltonian in the form
\be\label{eq:system-bath-expanded}
H_{SB} = \sum_{k=1}^\infty \sum_{\langle i_1,i_2, \dots i_k\rangle}H^{(k)}_{i_1,i_2, \dots i_k}
= \sum_{k=1}^\infty \frac{1}{k!}\sum_{i_1,i_2, \dots i_k}H^{(k)}_{i_1,i_2, \dots i_k}.
\ee
Here, $H^{(k)}_{i_1,i_2, \dots i_k}$ acts on the $k$ system qubits labeled by the indices $i_1,i_2, \dots i_k$, and also acts arbitrarily on the bath; for each $k$ we sum over all ways of choosing $k$ system qubits. We use $\langle i_1,i_2, \dots i_k\rangle$ to denote an unordered set of $k$ qubits; by definition, $H^{(k)}_{i_1,i_2, \dots i_k}$ is invariant under permutations of the $k $ qubits and vanishes if two of the indices coincide. Hence the two expressions for $H_{SB}$ in Eq.(\ref{eq:system-bath-expanded}) are equivalent. We will not need to assume anything about the initial state of the bath, except that the system qubits can be well enough isolated from the bath that we can prepare single-qubit states with reasonable fidelity.

We use the term {\em location} to speak of an operation in a quantum circuit performed in a single time step;
%where the duration of the step is at most $t_0$. 
a location may be a single-qubit or multi-qubit gate, a qubit preparation step, a qubit measurement, or the identity operation in the case where a qubit is idle during a time step. We model a noisy preparation as an ideal preparation followed by evolution governed by $H$, and a noisy measurement as an ideal measurement preceded by evolution governed by $H$. It is convenient to imagine that all system qubits are prepared at the very beginning of the computation and measured at the very end; in that case the noisy computation can be fully characterized by a unitary evolution operator $U$ acting jointly on the system and the bath, obtained by solving the time-dependent Schr\"odinger equation for the Hamiltonian Eq.(\ref{eq:system-bath-expanded}). 

%\noindent\emph{Explain justification of above assumptions.}

Let's briefly explain the physical justification for these assumptions. For fault-tolerant computing to work, there must be a mechanism for flushing the entropy introduced by noise; typically, entropy is removed from the computer by error-correction gadgets which use a supply of fresh ancilla qubits that are discarded after use. For mathematical convenience, we suppose that the initial state of the system includes all of the ancilla qubits that will be needed during the full course of the computation. To model the actual situation, in which ancilla qubits are prepared as needed just before being used, we also suppose that ancilla qubits are perfectly isolated from the bath until ``opened'' at the onset of the gadget in which they participate. Similarly, we suppose that the measurements of all ancilla qubits are delayed until the very end of the computation, but that these qubits are ``closed'' (their coupling to the bath is turned off) at the conclusion of the gadget in which they participate. Fault-tolerant gadgets sometimes also include quantum gates that are conditioned on the classical outcomes of earlier measurements. These conditional gates can be included in our framework; operations conditioned on measurements may be replaced by coherent gates, conditioned on the state of a ``closed'' control qubit that will be measured later (see Sec. VIC of \cite{NP09}). With these stipulations, our noise model is equivalent to a more realistic one in which ancilla qubits are repeatedly measured, reset, and reused. In this model we take for granted that ``pretty good'' fresh ancillas can be prepared at any time, or equivalently that qubits can be effectively erased at any time. A similar assumption would be needed to ensure scalability in an analysis of fault-tolerant reversible classical computation \cite{aharonov_reversible}.

%Implicitly, we have adopted a ``two-reservoir'' hypothesis. One reservoir, which we have called the ``bath,'' interacts with the system qubits, causing noise. The other reservoir is the entropy ``sink,'' which carries away heat each time a qubit is erased. In our model, the bath and the sink are uncoupled, and the sink has infinite heat capacity --- it never heats up no matter how many qubit erasures occur.

%Because the bath interacts with the system, in principle it might be driven far from its initial state in a manner that depends on the ideal computation being simulated. Our arguments have shown that, at least if the bath is a system of uncoupled oscillators and its initial state is Gaussian, the bath will not be pushed to a highly adversarial state that overpowers our efforts to make the computation robust. 
%One wonders how that conclusion could be altered if we relax the two-reservoir hypothesis by coupling the sink and the bath, or by eliminating the sink entirely. For example, we could attempt to model measurement and erasure more realistically by including entropy flow from the system to the bath. In that case, a bath of unbounded heat capacity would be needed to remove entropy from a noisy computation of unbounded size, and our modeling would need to incorporate a mechanism for equilibration of the bath. The goal would be to specify conditions under which the entropy flow from system to bath can be maintained well enough to support scalable quantum computation. For now, we put aside this ambitious project as an open problem for future consideration.

Our goal is to derive from Eq.(\ref{eq:system-bath-expanded}) an expression for the {\em effective noise strength} $\varepsilon$ of the noisy computation, which is defined as follows \cite{AGP06,AKP06}. We envision performing a formal expansion of $U$ in powers of the perturbation $H_{SB}$, to all orders. Consider a particular set ${\cal I}_r$ of $r$ circuit locations, and let $E({\cal I}_r)$ denote the sum of all terms in the expansion such that every location in ${\cal I}_r$ is faulty, {\em i.e.}, such that at least one of the qubits at that location is struck at least once by a term in $H_{SB}$ during the execution of the gate. We say that the noise has effective noise strength $\varepsilon$ if
\be
\| E({\cal I}_r) \| \le \varepsilon^r
\ee
for any set ${\cal I}_r$.
The accuracy threshold theorem for quantum computing shows that scalable quantum computing is possible if $\varepsilon$ is less than a positive constant $\varepsilon_0\approx 10^{-4}$ \cite{AGP06,NP09}. 

Let us define
\be\label{eq:eta-k-tilde}
\tilde \eta_1^{(k)}= \max_{i_1}\sum_{i_2, i_3,\dots, i_k} \|H^{(k)}_{i_1, i_2, i_3,\dots, i_k}\|~t_0,
\ee
where the maximum is over all system qubits and all times, the sum is over all system qubits, and $t_0$ is the maximal duration of any location. Then our main result can be stated as follows.

\begin{theorem}\emph{(Effective noise strength for correlated Hamiltonian noise)}
If each quantum gate acts on at most $m$ qubits and if 
\be
\tilde \eta_1^{(k)} \le f_k \alpha^k,
\ee
for all $k$, then
\be\label{eq:strength-theorem}
\varepsilon \le 2m \alpha~\exp\left(\sum_{k=1}^\infty \frac{g_k}{2k!} \right),
\ee
where
\be\label{eq:g-define}
g_k =  \sum_{l=0}^\infty\frac{(k-1)!f_{k+l}(2\alpha)^{l}}{(k+l-1)!}.
\ee
%{\em [No! Statement needs revision]}
\label{thm:effective-noise}
\end{theorem}

\noindent It follows that quantum computing is scalable provided the strength of $k$-qubit interactions decays sufficiently rapidly with $k$ (so that the sums in Eq.(\ref{eq:strength-theorem}) and Eq.(\ref{eq:g-define}) converge), and also decays as the spatial separation of the qubits increases (so that the sum defining $\tilde \eta_1^{(k)}$ in Eq.(\ref{eq:eta-k-tilde}) converges). 

If, for example, $f_k = 1$, then 
\be
g_k \le \sum_{l=0}^\infty (2\alpha)^l = \left(1 - 2\alpha\right)^{-1} \equiv C(\alpha),
\ee
and hence
\be
\varepsilon \le 2m\alpha\left(e^{(e-1)/2}\right)^{C(\alpha)}\approx 4.72 ~m\alpha,
\ee
where the last approximation uses $C(\alpha)\approx 1$ for $\alpha \ll 1$, as is the case if $\varepsilon$ is smaller than the threshold value $\varepsilon_0\approx 10^{-4}$. This observation can be restated as the following corollary:

\begin{corollary}%\emph{(Effective noise strength for correlated Hamiltonian noise)}
If each quantum gate acts on at most $m$ qubits then the effective noise strength can be expressed as 
\be
\varepsilon \le 2m\alpha\left(e^{(e-1)/2}\right)^{C(\alpha)},%\approx 4.72 ~m\alpha,
\ee
where 
\be
\alpha = \max_{k\ge 1}\left(\max_{i_1}\sum_{i_2, i_3,\dots, i_k} \|H^{(k)}_{i_1, i_2, i_3,\dots, i_k}\|~t_0\right)^{1/k},
\ee
$t_0$ is the maximal duration of any circuit location, and $C(\alpha) = \left(1-2\alpha\right)^{-1}$.
\label{coro:effective-noise1}
\end{corollary}

If instead
\be
f_k\le k!/k^p
\ee
where $p\ge 1$,
then
\be
g_k &\le& \frac{k!}{k^p}\left(\sum_{l=0}^\infty\frac{(k-1)!k^pf_{k+l}(2\alpha)^{l}}{(k+l-1)!k!}\right)=\frac{k!}{k^p}\left(\sum_{l=0}^\infty\frac{(k-1)!k^p(k+l)!(2\alpha)^{l}}{(k+l-1)!k!(k+l)^p}\right)\nonumber\\
&\le& \frac{k!}{k^p}\left(\sum_{l=0}^\infty\frac{k^{p-1}(2\alpha)^{l}}{(k+l)^{p-1}}\right)\le \frac{k!}{k^p}\left(\frac{1}{1-2\alpha}\right) =  \frac{k!}{k^p}C(\alpha).
\ee
For $p>1$ the sum over $k$ in Eq.(\ref{eq:strength-theorem}) converges, and hence we obtain a finite expression for $\varepsilon$; therefore, scalable fault-tolerant quantum computation is achievable for sufficiently small (nonzero) $\alpha$. We have obtained:
%%%

\begin{corollary}%\emph{(Effective noise strength for correlated Hamiltonian noise)}
If each quantum gate acts on at most $m$ qubits then for any $p > 1$ the effective noise strength can be expressed as 
\be
\varepsilon \le 2m \alpha_p~\exp\left(C(\alpha_p) \sum_{k=1}^\infty \frac{1}{2k^p} \right),
\ee
where
\be
\alpha_p = \max_{k\ge 1} \left[\frac{k^p}{k!}\left(\max_{i_1}\sum_{i_2, i_3,\dots, i_k} \|H^{(k)}_{i_1, i_2, i_3,\dots, i_k}\|~t_0\right)\right]^{1/k},
\ee
$t_0$ is the maximal duration of any circuit location, and $C(\alpha) = \left(1-2\alpha\right)^{-1}$.
\label{coro:effective-noise2}
\end{corollary}

%%%
\noindent In particular, if $p=2$ for example, we find
\be
\varepsilon \le 2m\alpha \left(e^{\pi^2/12}\right)^{C(\alpha)}\approx 4.55~m \alpha,
\ee
again using $C(\alpha) \approx 1$ to obtain the numerical expression. Using either Corollary 1 or Corollary 2, together with the results in \cite{AGP06,NP09}, we conclude that fault-tolerant quantum computing is scalable provided that $\alpha$ is less than a constant $\alpha_0 \approx 10^{-5}$. 

\section{Proof of Theorem 1}
\label{sec:proof}

In \cite{AKP06}, scalability was proven for the special case in which only the $k=2$ term in the Hamiltonian is nonzero. To prove Theorem \ref{thm:effective-noise} we generalize the ideas used in \cite{AKP06}. We write the system-bath Hamiltonian as
\be
H_{SB} = \sum_a H_{SB,a}
\ee
where $a$ is a shorthand for the indices $k$, and $i_1,i_2, \dots i_k$ in Eq.(\ref{eq:system-bath-expanded}). For the sake of conceptual clarity we imagine dividing time into infinitesimal intervals, each of width $\Delta$, and express the time evolution operator for the interval $(t,t+\Delta)$ as
\be\label{eq:U-delta}
U(t+\Delta, t) \approx e^{-i\Delta H} \approx e^{-i\Delta H_S}e^{-i\Delta H_B}\prod_a(I_{SB}-i\Delta H_{SB,a}). 
\ee
(We have omitted terms higher order in $\|H\| \Delta$; strictly speaking, then, to justify Eq.(\ref{eq:U-delta}) we should regulate the bath Hamiltonian $H_B$ by imposing an upper bound on its norm, then choose $\Delta$ small enough so these higher order terms can be safely neglected.) We expand $U(t+\Delta,t)$ as a sum of monomials, where for each value of $a$ either $I_{SB}$ or $-i\Delta H_{SB,a}$ appears; then we obtain the perturbation expansion of the full time evolution operator $U$ over time $T$ by stitching together $T/\Delta$ such infinitesimal time evolution operators. 

%\begin{figure}[tbh]
\begin{figure}[t]
\begin{center}
\includegraphics[width=0.6\textwidth]{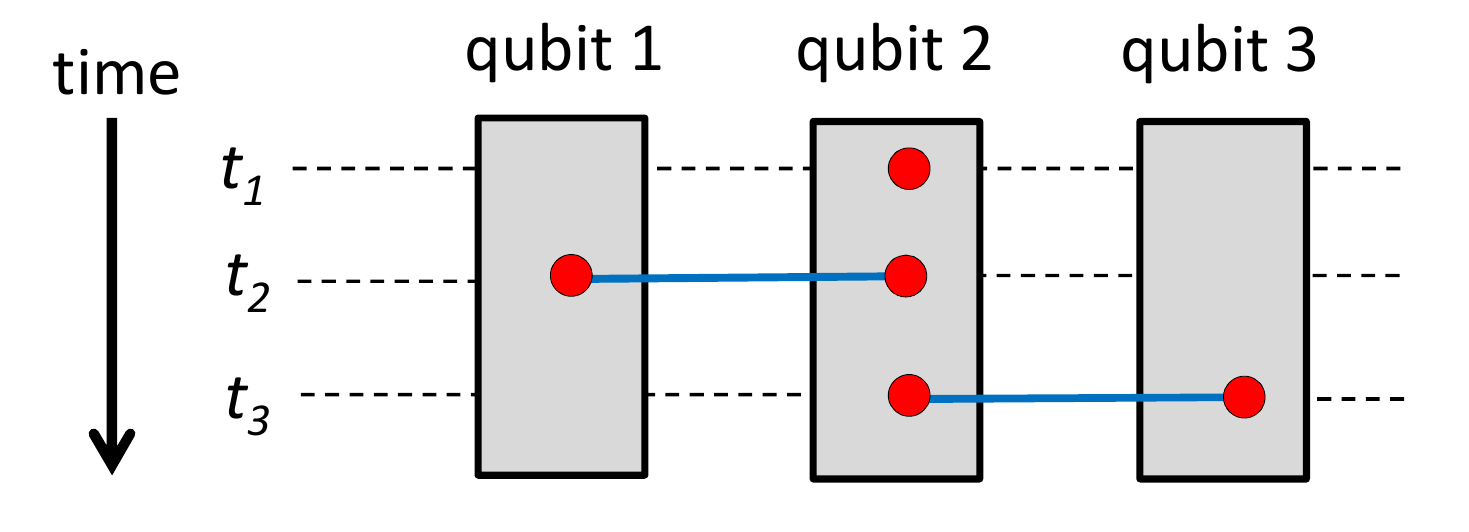}
\end{center}
\fcaption{\label{fig1} Three faulty single-qubit locations in a single time step, with time flowing vertically downward. The first insertions of the system-bath Hamiltonian $H_{SB}$ at these three locations are indicated: $H^{(1)}_2$ acts on qubit 2 at time $t_1$, $H^{(2)}_{12}$ acts on qubits 1 and  2 at time $t_2$, and $H^{(2)}_{23}$ acts on qubits 2 and 3 at time $t_3$. In the ``modified Hamiltonian'' $H_{SB}^{\rm (modified)}$, the term $H^{(1)}_2$ is turned off during the portion of this time step prior to $t_1$, $H^{(2)}_{12}$ is turned off prior to $t_2$, and $H^{(2)}_{23}$ is turned off prior to $t_3$. }
\end{figure}

We will refer to the $r$ specified locations in the set ${\cal I}_r$ as the ``marked locations'' and to the remaining locations as the ``unmarked locations.'' For now, suppose for definiteness that all of the marked locations are single-qubit gates. For any term in the perturbation expansion contributing to $E({\cal I}_r)$ there must be an {\em earliest} infinitesimal time interval in each of the $r$ marked locations where a term $H_{SB,a}$ acts nontrivially on that qubit. Suppose we fix the infinitesimal time intervals where these earliest ``insertions'' of $H_{SB}$ occur, and also fix the terms $\{H_{SB,a}\}$ in the system-bath Hamiltonian that act there, but sum over all the terms in the perturbation expansion acting in other infinitesimal time intervals and on other qubits. Consider a particular step in the computation executed during the time interval $[t,t+T]$, and a pair of successive fixed earliest insertions of $H_{SB}$ occurring at times $t_1, t_2$, where $t < t_1 < t_2 < t+T$. Then during the time interval $[t_1, t_2]$ the joint evolution of the system and the bath is governed by a modified Hamiltonian
\be
H^{\rm (modified)}([t_1,t_2])& =& H_S +H_B +\sum_a H_{SB,a}^{\rm (modified)}([t_1,t_2]);
\ee
here, $H^{(k){\rm (modified)}}_{i_1, \dots, i_k}$ is set to zero if a first insertion of $H_{SB}$ acts on any of the qubits $i_1, i_2, \dots i_k$ at time $s \in [t_2,t+T]$; otherwise, $H^{(k){\rm (modified)}}_{i_1, \dots, i_k}= H^{(k)}_{i_1, \dots, i_k}$. (See Fig. 1.) The important point is that the time evolution operator in between successive insertions of the perturbation is unitary and hence has unit operator norm. Using the submultiplicative property of the operator norm, then, we conclude that the contribution to $E({\cal I}_r)$ with the earliest insertions at the marked locations fixed has operator norm bounded above by
\be
\prod_a^{\rm (earliest)}\left( \|H_{SB,a} \|\Delta\right),
\ee
where the product is over the terms in the system-bath Hamiltonian that act at the earliest insertions. To bound $\|E({\cal I}_r)\|$, we sum over the $t_0/\Delta$ time intervals at each location where the earliest insertion may occur, and also sum over all the ways of choosing the term $H_{SB,a}$ that acts at each insertion, obtaining
\be
\|E({\cal I}_r)\| \le \sum_{\{H_{SB,a}\}}^{\rm (insertions)} \prod_a^{\rm (earliest)}\left( \|H_{SB,a} \| t_0\right).
\ee
Summing over the possible intervals for the first insertion turns the factor $\Delta$ into the factor $t_0$.

Now we have to figure out how to sum over all ways of choosing the terms $\{H_{SB,a}\}$ acting at the earliest insertions inside the $r$ marked circuit locations. Since $H_{SB}$ contains multi-qubit terms, a single term in $H_{SB}$ can simultaneously produce the first insertion at multiple circuit locations occurring in the same time step. Specifically, a single term in $H^{(k)}$ might cause simultaneous faults in $j$ of the $r$ marked locations for any $j\le k$, if all of these $j$ locations occur in the same time step. We use the term ``$j$-contraction'' to refer to the case where a single term in Eq.(\ref{eq:system-bath-expanded}) produces the first insertion in each of $j$ marked locations. (See Fig. 2.)

%\begin{figure}[tbh]
\begin{figure}[t]
\begin{center}
\includegraphics[width=0.8\textwidth]{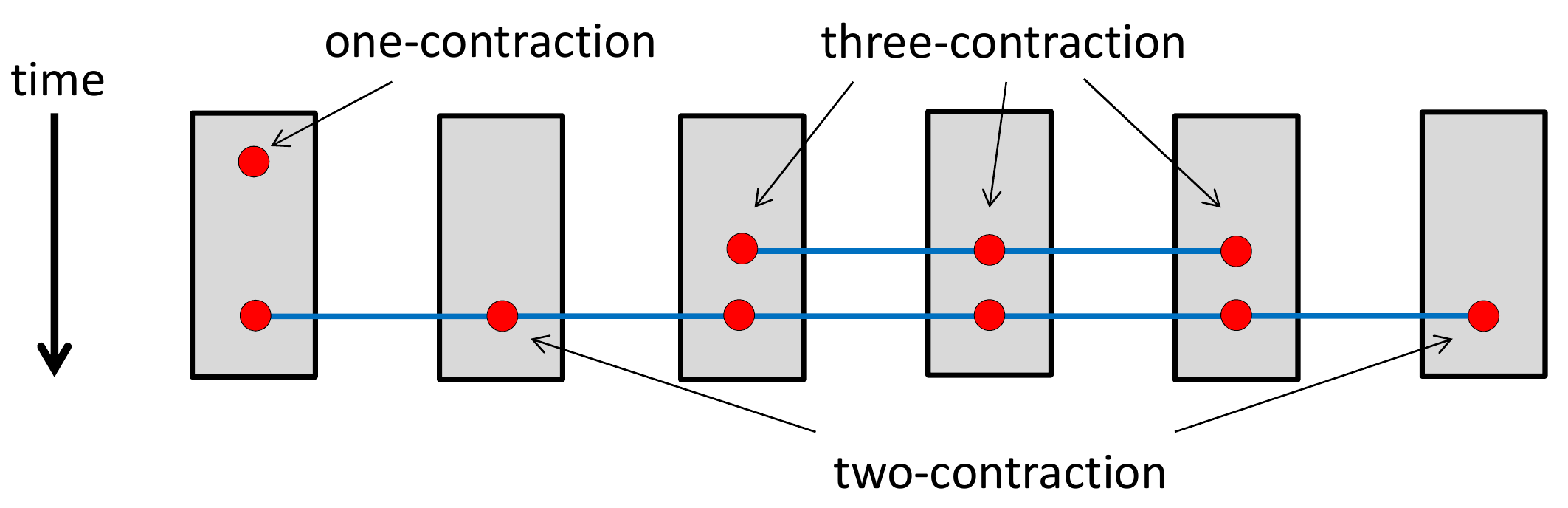}
\end{center}
\fcaption{\label{fig2} Contractions occurring during one computational time step, with time flowing vertically downward, where rectangles represent qubits and $k$ dots connected by a horizontal line signify an insertion of a term in $H^{(k)}$, which acts on $k$ qubits. All six of the qubits shown have faults during this time step, arising from a one-contraction, a three-contraction, and a two-contraction. The two-contraction is due to a term in $H^{(6)}$ that actually afflicts all six qubits, but counts as a two-contraction because four of these qubits have already been hit by other contractions earlier in the same time step.}
\end{figure}

First we find an upper bound on the strength of a one-contraction, the operator norm of the sum of all terms that cause one particular circuit location to be faulty. If the qubit at the marked location carries the label $i_1$, the term in the Hamiltonian responsible for the earliest insertion at this location could be any $H^{(1+l)}_{i_1, j_1, j_2,\dots, j_l}$ for $l\ge 0$; here for each $m=1, 2, \dots, l$ either qubit $j_m$ is unmarked or else qubit $j_m$ is marked but has already been struck by an earlier insertion during the same time step. Hence an upper bound on the strength of the one-contraction is
\be\label{eq:eta-one-bound}
\eta_1 = \sum_{l=0}^\infty \eta_1^{(1+l)},
\ee
where 
\be\label{eq:eta-one-ell}
\eta_1^{(1+l)} =\max^{\rm (in)}_{i_1}\sum^{\rm (all)}_{ \langle j_1, j_2,\dots, j_l\rangle } \|H^{(1+l)}_{i_1, j_1, j_2,\dots, j_l}\|~t_0
= \max^{\rm (in)}_{i_1}\frac{1}{l!}\sum^{\rm (all)}_{ j_1, j_2,\dots, j_l } \|H^{(1+l)}_{i_1, j_1, j_2,\dots, j_l}\|~t_0.
\ee
Here, to obtain an upper bound, we sum each index $j_m$ over all qubits, whether marked or unmarked, and also obtain the factor $t_0$ by summing over all the infinitesimal time intervals during a single time step. 
%all locations (both marked and unmarked), to allow for the possibility that the insertion at a marked location need not be the {\em first} insertion; hence the higher order ($l> 0$) terms in Eq.(\ref{eq:eta-one-ell}) can be contributions to the strength of the one-contraction rather than a $j$-contraction for $j>1$ even though some of the locations in $\{j_1,j_2, \dots, j_l\}$ may be marked. 
We have also maximized this expression over all possible ways to choose qubit $i_1$ from among the marked qubits --- the superscript ``in'' in $\max^{\rm (in)}$ indicates that we maximize over only the marked qubits, while the superscript ``all'' in $\sum^{\rm (all)}$ indicates that we sum over all qubits without any restriction. Of course, our upper bound would still be valid were we to relax the restriction and maximize over all qubits, whether marked or not.

Similarly, for $k>1$, the strength of a $k$-contraction can be bounded by
\be
\eta_k = \sum_{l=0}^\infty \eta_k^{(k+l)},
\ee
where
\be
\eta_k^{(k+l)} = \sum^{\rm (in)}_{\langle i_1, i_2, \dots, i_k\rangle }\sum^{\rm (all)}_{\langle j_1, j_2,\dots, j_l \rangle} \|H^{(k+l)}_{i_1, i_2, \dots i_k, j_1, j_2,\dots, j_l}\|~t_0.
\ee
Here, for the ``in'' sum the qubits are restricted to the marked locations and for the ``all'' sum they may be at either marked or unmarked locations. Note that the upper bound $\eta_1$ involves a maximum over marked qubits, while the upper bound $\eta_k$ for $k > 1$ involves instead a sum over marked qubits; the reason for this distinction is explained in the next paragraph. 

By summing all ways of choosing the first insertion in each of $r$ marked locations, we obtain the bound
\be\label{eq:contraction-sum}
\varepsilon^r \le  \sum^{(r)}_{r_1, r_2, r_3, \dots}\prod_{k=1}^\infty \frac{1}{r_k!}\left(\eta_k\right)^{r_k}.
\ee
Here $r_k$ is the number of $k$-contractions, and the sum $\sum^{(r)}$ is subject to the constraint $\sum_k k r_k = r$. To obtain Eq.(\ref{eq:contraction-sum}), we observe that, for $k > 1$,
\be
\left( \sum_{l=0}^\infty \sum^{\rm (in)}_{\langle i_1, i_2, \dots, i_k\rangle }\sum^{\rm (all)}_{\langle j_1, j_2,\dots, j_l \rangle} \|H^{(k+l)}_{i_1, i_2, \dots i_k, j_1, j_2,\dots, j_l}\|~t_0\right)^{r_k}
\ee
contains each way of choosing $r_k$ $k$-contractions among the $r$ marked locations $r_k!$ times, plus additional nonnegative terms; the factor $1/r_k!$ in Eq.(\ref{eq:contraction-sum}) compensates for this overcounting. Furthermore, once all the higher rank contractions have been fixed, the locations where one-contractions occur are completely determined. That is why we defined $\eta_1$ by maximizing over $i_1$, rather than summing $i_1$ over all the marked qubits.

%[\emph{Explain more?}]

In Eq.(\ref{eq:eta-one-bound}) we have derived a bound on the strength of the noise acting at a single circuit location. We wish to go further and investigate whether the correlations in noise acting collectively on many circuit locations could overcome fault-tolerant protocols, even if the individual gates perform very well. For this purpose, we should relate $\eta_k$ for $k> 1$ to $\eta_1$. Note that in $\eta_k^{(k+l)}$ for $k>1$, we can replace the sum over ways to choose $k$ qubits by a sum over all qubits divided by $k!$, and similarly we can replace the sum over the ways to choose $l$ qubits by a sum over all qubits divided by $l!$, obtaining
\be
\eta_k^{(k+l)} = \frac{1}{k!}\sum^{\rm (in)}_{ i_1, i_2, \dots, i_k }\frac{1}{l!}\sum^{\rm (all)}_{ j_1, j_2,\dots, j_l } \|H^{(k+l)}_{i_1, i_2, \dots i_k, j_1, j_2,\dots, j_l}\|~t_0.
\ee 
By summing $i_1$ over the $r$ marked locations we obtain the bound
\be
\eta_k^{(k+l)} \le r~\max_{i_1}^{\rm (in)} \frac{1}{k!}\sum^{\rm (all)}_{ i_2, \dots, i_k }\frac{1}{l!}\sum^{\rm (all)}_{ j_1, j_2,\dots, j_l } \|H^{(k+l)}_{i_1, i_2, \dots i_k, j_1, j_2,\dots, j_l}\|~t_0;
\ee 
note that we still have a bound if we extend the ``in'' sum to a sum over all qubits.
From Eq.(\ref{eq:eta-one-ell}) we have
\be
\eta_1^{(k+l)}= \max^{\rm (in)}_{i_1}\frac{1}{(k+l-1)!}\sum^{\rm (all)}_{ i_2, \dots, i_k }\sum^{\rm (all)}_{ j_1, j_2,\dots, j_l } \|H^{(k+l)}_{i_1, i_2, \dots i_k, j_1, j_2,\dots, j_l}\|~t_0.
\ee
which implies, for $k> 1$,
\be
\eta_k^{(k+l)} \le r \frac{(k+l-1)!}{k!~l!} \eta_1^{(k+l)}= \frac{r}{k} %\binom{k+l -1}{l}  
{k+l-1 \choose l} \eta_1^{(k+l)}\le \frac{r}{k}2^{k+l-1} \eta_1^{(k+l)}.
\ee
Hence we find, for $k> 1$,
\be\label{eq:eta-k-series}
\eta_k \le \frac{r}{2k}\sum_{l=0}^\infty 2^{k+l} \eta_1^{(k+l)}.
\ee
This is the key inequality that we needed, relating (an upper bound on) the strength of collective noise acting on $k$ circuit locations to a sum over (upper bounds on) contributions to the noise strength for a single location.

Now suppose, as in the hypothesis of Theorem 1, that
\be
\eta_1^{(k)}=  \frac{1}{(k-1)!}~\tilde \eta_1^{(k)} \le \frac{f_k \alpha^k}{(k-1)!}.
\ee
%The reason to include the $k!$ is that we are supposing that the $\sim \alpha^k$ bound holds when we sum all locations independently rather than summing over sets of qubits. 
From Eq.(\ref{eq:eta-k-series}) we obtain
\be
\eta_k \le \frac{rg_k}{2k!}(2\alpha)^k,
\ee
where
\be
g_k = f_k + \sum_{l=1}^\infty\frac{(k-1)!f_{k+l}(2\alpha)^{l}}{(k+l-1)!}.
\ee
Then the bound Eq.(\ref{eq:contraction-sum}) becomes
\be
\varepsilon^r \le \sum^{(r)}_{r_1, r_2, r_3, \dots}\prod_{k=1}^\infty \frac{1}{r_k!}\left(\frac{rg_k(2\alpha)^k}{2k!}\right)^{r_k}
= (2\alpha)^r \sum^{(r)}_{r_1, r_2, r_3, \dots}\prod_{k=1}^\infty \frac{1}{r_k!}\left(\frac{rg_k}{2k!} \right)^{r_k},
\ee
recalling the constraint on the sum.
If we now relax the constraint on the sum, we have
\be
\varepsilon^r &\le& \left(2\alpha\right)^r\prod_{k=1}^\infty \sum_{r_k =0}^\infty\frac{1}{r_k!}\left(\frac{rg_k}{2k!} \right)^{r_k}=\left(2\alpha\right)^r\prod_{k=1}^\infty\exp\left(\frac{rg_k}{2k!}\right) \nonumber\\
&=& \left(2\alpha\right)^r\left(\exp\left(\sum_{k=1}^\infty \frac{g_k}{2k!} \right)\right)^r = \left(2 \alpha~\exp\left(\sum_{k=1}^\infty \frac{g_k}{2k!} \right) \right)^r.
\ee

Up until now, we have considered all circuit locations to be single-qubit locations. In the case of an $m$-qubit gate location, the location is faulty if the system-bath perturbation acts nontrivially on any one of $m$ qubits, which enhances each $\eta_k$ appearing in Eq.(\ref{eq:contraction-sum}) by at most a factor of $m^k$, and hence increases our upper bound on the effective noise strength by at most a factor of $m$. This completes the proof of Theorem 1. 
\hfill $\square$

\section{Conclusions and outlook}
\label{sec:conclusions}
Theorem 1, combined with results from \cite{AGP06}, shows that fault-tolerant quantum computing is scalable in principle for a class of correlated noise models. The key feature of these models is that, while there are terms in the noise Hamiltonian acting collectively on $k$ system qubits for $k \gg 1$ (and simultaneously on the quantum computer's environment, {\em i.e.}, the ``bath''), the operator norm of these terms decays faster than any power of $k$. The result may apply even if, for each fixed $k$, the operator norm of the $k$-qubit noise term decays algebraically rather than exponentially as the qubits are spatially separated, provided the decay is sufficiently rapid as a function of distance for the sum over system qubits in Eq.(\ref{eq:eta-k-tilde}) to converge. 

Theorem 1 generalizes results found in \cite{AKP06,Novais07} for the case $k=2$. Though the analysis is formulated in terms of interaction-picture perturbation theory, it is rigorous because our estimate of the effective noise strength is derived by bounding perturbation theory summed to all orders.

We also assume that at all times during the computation it is possible to prepare qubits in a standard initial state and to measure qubits in a standard basis with reasonable fidelity, but we make no other assumptions about the state of the bath. Although the bath might be quite ``hot,'' the flow of entropy from the bath to the computer is impeded by the weak coupling between the system and the bath, allowing a fault-tolerant protocol to maintain a steady state where entropy is removed fast enough to keep the computer ``cool.'' Entropy is carried away by the ancilla qubits used in error correction gadgets, which are subsequently erased and reused. A mechanism for flushing entropy is required in any scheme for stabilizing a noisy computer, whether quantum or classical \cite{aharonov_reversible}.

Of course, the condition for scalability derived in Theorem 1 is merely sufficient and not necessary. In particular, scalability may be provable even if the system-bath Hamiltonian has unbounded norm, but in that case further assumptions are needed about the state of the bath at the beginning of the computation. For example, threshold theorems for Gaussian noise were proven in \cite{NP09}, which apply if the bath is a linear system of harmonic oscillators and the initial state of the bath is Gaussian (a thermal state for example). In that case, the criterion for scalability can be stated as a property of the bath's two-point correlation function, ensuring that spatial and temporal noise correlations decay sufficiently rapidly. In this particular setting, at least, we can address the concern expressed in \cite{AHHH02} that over the course of a long computation an oscillator bath could be driven to a highly adversarial state that overwhelms fault-tolerant protocols.

The purpose of this work is to address an issue of principle concerning the scalability of quantum computers subject to correlated noise, not to obtain optimized realistic estimates of the accuracy threshold for quantum computing. Indeed the noise strength $\varepsilon$ appearing in our criterion for scalability is in effect an error amplitude per gate rather than an error probability per gate, and must be below about $10^{-4}$ for known threshold theorems to apply $\cite{AGP06,NP09,Terhal05}$. This criterion is probably much too pessimistic --- for the general class of Hamiltonian noise models considered here, we cannot easily rule out substantial enhancement of the logical failure probability due to constructive interference of many coherently combined fault histories, even though this seems quite unlikely in practice. Furthermore, we have not attempted here to assess the effectiveness of methods such as noiseless subsystems \cite{Knill00} or dynamical decoupling \cite{Viola99} which could suppress the noise correlations. Characterizing the residual noise correlations when dynamical decoupling is employed may be difficult for general Hamiltonian noise models, though some preliminary steps were reported in \cite{NLP11}. Also, to derive Theorem 1 we made no assumptions about the bath Hamiltonian $H_B$, and it may be possible to derive stronger results contingent on physically motivated limitations on the bath dynamics such as locality constraints. 

The modest results derived here can hardly be expected to assuage the quantum computing skeptics, but may nevertheless help to clarify the debate. 
%In a world in the midst of a digital revolution, 
Can we identify fundamental principles of physics that are compatible with large-scale classical computing yet incompatible with large-scale quantum computing? Enlarging the class of noise models for which quantum computing is provably scalable should shed light on this fascinating and important question. If quantum mechanics is valid and if future quantum engineers can devise a controllable many-qubit system with noise meeting the criterion derived in this paper, then reliable large-scale quantum computing will be achievable.

%Physics is local.
%What if the correlations decay algebraically? Would it help if the bath has an energy gap?
%Many intriguing questions remain open. 
%What models of noise are classically but not quantumly scalable?
%What deformations from QM might still allow superclassical computation. 
%A challenge for the quantum engineers

\nonumsection{Acknowledgments}
\noindent
I thank Dick Lipton and Ken Regan for allowing me to post a link to a preliminary account of this work on their blog \emph{G\"odel's Lost Letter}, and I thank the many readers who posted useful comments on the blog, especially Robert Alicki, Joe Fitzsimons, Aram Harrow, Gil Kalai, and John Sidles. I also thank Peter Brooks, Michael Beverland, Daniel Lidar, and Gerardo Paz-Silva for discussions. 
%%%
This work was supported in part by the Intelligence Advanced Research Projects Activity (IARPA) via Department of Interior National Business Center contract number D11PC20165. The U.S. Government is authorized to reproduce and distribute reprints for Governmental purposes notwithstanding any copyright annotation thereon. The views and conclusions contained herein are those of the author and should not be interpreted as necessarily representing the official policies or endorsements, either expressed or implied, of IARPA, DoI/NBC
or the U.S. Government.
%%%
I also acknowledge support from NSF grant PHY-0803371, DOE grant DE-FG03-92-ER40701, and NSA/ARO grant W911NF-09-1-0442. The Institute for Quantum Information and Matter (IQIM) is an NSF Physics Frontiers Center with support from the Gordon and Betty Moore Foundation. 

\newpage
\nonumsection{References}

\end{document}